\documentstyle[bo99,epsfig]{article}
\title{Oscillations During Thermonuclear X-ray Bursts: A New Probe of 
Neutron Stars }
\author{Tod E. Strohmayer }
\affil{NASA's Goddard Space Flight Center, Greenbelt, MD 20771 }

\begin{document}

\maketitle

\begin{abstract}

Observations of thermonuclear (also called Type I) X-ray bursts from neutron stars
in low mass X-ray binaries (LMXB) with the Rossi X-ray Timing Explorer (RXTE) 
have revealed large amplitude, high coherence X-ray brightness oscillations with 
frequencies in the 300 - 600 Hz range. Substantial spectral and timing evidence 
point to rotational modulation of the X-ray burst flux as the cause of these 
oscillations, and it is likely that they reveal the spin frequencies of neutron 
stars in LMXB from which they are detected.  Here I review the status of our 
knowledge of these oscillations and describe how they can be used to constrain 
the masses and radii of neutron stars as well as the physics of thermonuclear 
burning on accreting neutron stars. 

\end{abstract}

\section{Introduction }

During the past 25 years X-ray astronomers have expended a good deal of effort in
an attempt to identify the spin frequencies of neutron stars in LMXB (see for
example Wood et al. 1991; Vaughan et al. 1994). These efforts were largely 
prompted by the discovery in the radio band of rapidly rotating neutron 
stars, the millisecond radio pulsars (see Backer et al. 1984), and subsequent 
theoretical work suggesting their origin lie in an accretion-induced spin-up 
phase of LMXB (see the review by Bhattacharya 1995 and references therein). 
However, up to the mid-90's there was little or no direct evidence to support 
the existence of rapidly spinning neutron stars in LMXB. This situation 
changed dramatically with the launch of the {\it Rossi X-ray Timing Explorer} (RXTE) 
in December, 1995. Within a few months of its launch RXTE observations had
provided strong evidence suggesting that neutron stars in LMXB are spinning
with frequencies $\ge 300$ Hz. These first indications came with the discovery
of high frequency (millisecond) X-ray brightness oscillations, ``burst oscillations,''
during thermonuclear (Type I) X-ray bursts from several neutron star LMXB systems 
(see Strohmayer et al. 1996; Smith, Morgan \& Bradt 1997; Zhang et al. 1996). 

At present these oscillations have been observed from six different LMXB systems 
(see Strohmayer, Swank \& Zhang 1998). The observed frequencies are in the range 
from $\approx 300 - 600$ Hz, similar to the observed frequency distribution of 
binary millisecond radio pulsars (Taylor, Manchester \& Lyne 1993), and consistent 
with some theoretical determinations of spin periods which can be reached via 
accretion-induced spin-up (Webbink, Rappaport \& Savonije 1983). In this contribution 
I will review our observational understanding of
these oscillations, with emphasis on how they can be understood in the context of
spin modulation of the X-ray burst flux. I will discuss how detailed modelling of
the oscillation amplitudes and harmonic structure can be used to place interesting
constraints on the masses and radii of neutron stars and therefore the equation of
state of supranuclear density matter. Inferences which can be drawn regarding the 
physics of thermonuclear burning will also be discussed. I will conclude with some 
outstanding theoretical questions and uncertainties and where future observations 
and theoretical work may lead.

\section{Observational properties of burst oscillations}

Burst oscillations with a frequency of 363 Hz were first discovered from the LMXB 
4U 1728-34 by Strohmayer et al. (1996). Since then an additional five sources 
with burst oscillations have been discoverd. The burst oscillation sources and
their observed frequencies are given in table 1. In the remainder of this section
I review the important observational properties of these oscillations and 
attempt to lay out the evidence supporting the spin modulation hypothesis.

\subsection{Oscillations at burst onset}

Many bursts show detectable oscillations during the $\approx 1 - 2$ s risetimes 
typical of thermonuclear bursts. For example, Strohmayer, Zhang \& Swank (1997) 
showed that some bursts from 4U 1728-34 have oscillation amplitudes as large as
43 \% within 0.1 s of the observed onset of the burst. They also showed that the
oscillation amplitude decreased monotonically as the burst flux increased during
the rising portion of the the burst lightcurve.  Figure 1 shows this behavior
in a burst from 4U 1636-53. This burst had an oscillation amplitude near onset of
$\approx 80$ \%, and then showed an episode of radius expansion beginning near
the time when the oscillation became undectable (see Strohmayer et al. 1998a).
The presence of modulations of the thermal burst flux approaching 100 \% right
at burst onset fits nicely with the idea that early in the burst there exists a
localized hot spot which is then modulated by the spin of the neutron star. In 
this scenario the largest modulation amplitudes are produced when the spot
is smallest, as the spot grows to encompass more of the neutron star surface, the
amplitude drops, consistent with the observations. 

\begin{figure}
\centerline{\psfig{file=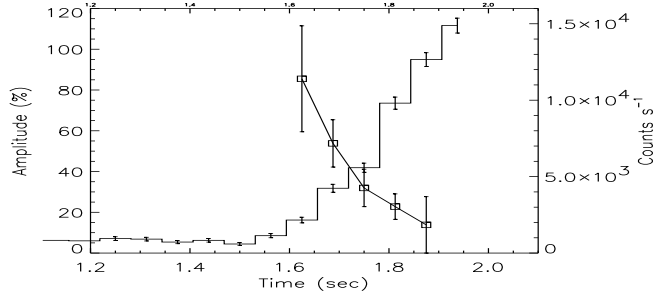, height=4cm,width=9cm,clip=}}
\vskip 10pt
\caption{The amplitude of oscillations at 580 Hz during the rising
phase of a burst from 4U 636-53. The amplitude is largest near the 
onset of the burst and decreases as the burst flux increases.}
\end{figure}

X-ray spectroscopy during burst rise also suggests that the emission is localized
near the onset of bursts. Prior to RXTE few instruments had the collecting area and 
temporal resolution to study spectral evolution during the short rise times of
thermonuclear bursts. Day \& Tawara (1990) used GINGA
observations of 4U 1728-34 in an attempt to constrain the e-folding spreading 
time of the burning front to $\approx 0.1$ s in two bursts. With RXTE Strohmayer, 
Zhang \& Swank (1997) investigated the spectral evolution of bursts 
from 4U 1728-34. They fit a black body model to intervals during several bursts  
and plotted the flux $F_{bol}$ versus $F_{bol}^{1/4} / kT_{BB}$. For black body 
emission from a spherical surface this ratio is a constant proportional to 
$(R/d)^{1/2}$, where $d$ and $R$ are the source distance and radius, respectively. 
Figure 2 shows such a plot for a burst from 4U 1728-34. In this plot 
the solid line connects successive time intervals, with the burst beginning in the
lower left and evolving diagonally to the upper right and then across to the left.
This evolution indicates that the X-ray emitting area is {\it not} constant, but 
{\it increases} with time during the burst rise. The spectra of type I bursts are 
not true black bodies (see London, Taam \& Howard 1986; Ebisuzaki 1987; Lewin, 
van Paradijs \& Taam 1993), however, the argument here concerns the energetics and 
not the detailed shape of the spectrum. Since the effect is seen in  bursts that 
do not show photospheric radius expansion the atmosphere is always geometrically 
thin compared with the stellar radius, so the physics of spectral formation 
depends only on conditions locally. The {\it same} spectral hardening corrections 
will apply to intervals during the rising and cooling portions of the burst 
which have the same measured black body temperatures (ie. spectral shapes), 
thus, the deficit in $F_{bol}^{1/4} / kT_{BB}$ evident during the rising phase
cannot be due solely to spectral formation effects. 

\begin{table}
\caption{Burst oscillation sources and frequencies}
\centerline{\begin{tabular}{cc}
Object & Frequency (Hz) \cr
4U 1728-34 & 363 \cr
4U 1636-53 & 580 (290) \cr
4U 1702-429 & 330 \cr 
KS 1731-26 & 526 \cr
Aql X-1 & 549 \cr
Gal. Center & 589 \cr
\end{tabular}}
\end{table}

\subsection{Expectations from the theory of thermonuclear burning}

The thermonuclear instability which triggers an X-ray burst burns in a few 
seconds the fuel which has been accumulated on the surface 
over several hours. This $>$ 10$^3$ difference between the accumulation and 
burning timescales means that it is unlikely that the conditions 
required to trigger the instability will be achieved simultaneously over the 
entire stellar surface. This realization, first emphasized by Joss (1978), 
led to the study of lateral propagation of the burning over the neutron star
surface (see Fryxell \& Woosley 1982, Nozakura, Ikeuchi \& Fujimoto 1984, 
and Bildsten 1995). The subsecond risetimes of thermonuclear X-ray bursts 
suggests that convection plays an important role in the physics of the burning 
front propagation, especially in the low accretion rate regime which leads to 
large ignition columns (see Bildsten (1998) for a review of
thermonuclear burning on neutron stars). Bildsten (1995) has shown
that pure helium burning on neutron star surfaces is in general inhomogeneous,
displaying a range of behavior which depends on the local accretion rate, with
low accretion rates leading to convectively combustible accretion columns and 
standard type I bursts, while high accretion rates lead to slower, nonconvective 
propagation which may be manifested in hour long flares. These studies emphasize 
that the physics of thermonuclear burning is necessarily a multi-dimensional 
problem and that {\it localized} burning is to be expected, especially at the 
onset of bursts.  The propertiess of oscillations near burst onset described
above fit well into this picture of thermonuclear burning on neutron stars.

Miller (1999) has recently found evidence for a significant 290 Hz subharmonic
of the strong 580 Hz oscillation seen in bursts from 4U 1636-53. Evidence
for the subharmonic was found by adding together in phase data from the rising
intervals of 5 bursts. This result suggests that in 4U 1636-53 the spin 
frequency is 290 Hz, and that the strong signal at 580 Hz is caused by nearly
antipodal hot spots. If correct this result has interesting implications
for the physics of nuclear burning, in particular, how the burning 
can be spread from one pole to the other within a few tenths of seconds, and
how fuel is pooled at the poles, perhaps by a magnetic field. 

\begin{figure}[htb]
\centerline{\psfig{file=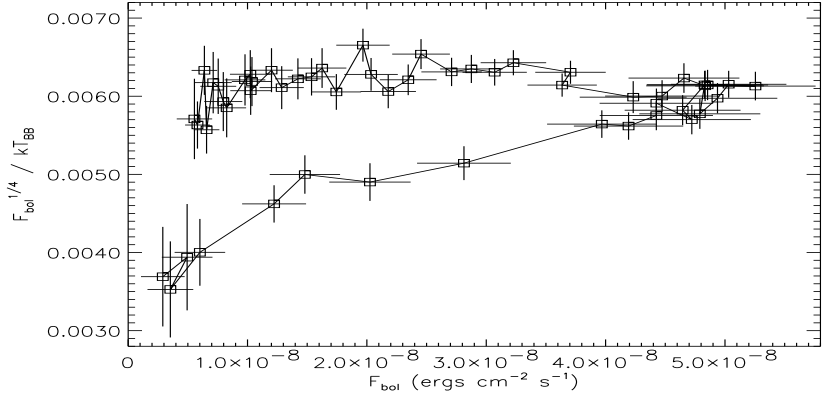,width=4.5in,height=2.25in,clip=}}
\caption{Bolometric flux versus $F_{bol}^{1/4} /kT_{BB}$ for a burst from 
4U 1728-34. The burst evolves from lower left to upper right and then to the left,
and is strong evidence for an increasing X-ray emission area during the burst rise.}
\end{figure}

\subsection{The coherence of burst oscillations}

One of the most interesting aspects of the burst oscillations is the 
frequency evolution evident in many bursts. The frequency
is observed to increase by $\approx 1 - 3$ Hz in the cooling tail, 
reaching a plateau or asymptotic limit (see Strohmayer et al. 1998a). 
An example of this behavior in a burst from 4U 1702-429 is shown in 
figure 3. However, increases in the oscillation frequency 
are not universal, Strohmayer (1999) and Miller (1999) have recently 
reported on an episode of {\it spin down} in the cooling tail of a burst from 
4U 1636-53. Frequency evolution has been seen in five of the six burst 
oscillation sources and appears to be commonly associated with the physics 
of the modulations. Strohmayer et. al (1997) have argued this evolution results 
from angular momentum conservation of the thermonuclear shell. The burst 
expands the shell, increasing its rotational moment of inertia and slowing 
its spin rate. Near burst onset the shell is thickest and thus the observed 
frequency lowest. The shell then spins back up as it cools and recouples to the
bulk of the neutron star. Calculations indicate that the $\sim 10$ m
thick pre-burst shell expands to $\sim 30$ m during the flash 
(see Joss 1978; Bildsten 1995), which gives a frequency shift due to
angular momentum conservation of $\approx 2 \ \nu_{spin} (20 \ {\rm m}/ R)$,
where $\nu_{spin}$ and $R$ are the stellar spin frequency and radius, 
respectively. For the several hundred Hz spin frequencies inferred from 
burst oscillations this gives a shift of $\sim 2$ Hz, similar to that observed.

In bursts where frequency drift is evident the drift broadens the peak in the
power spectrum, producing quality factors $Q \equiv \nu_0 / \Delta\nu_{FWHM}
\approx 300$. In some bursts a relatively short train of pulses is observed 
during which there is no strong evidence for a varying frequency. Recently,
Strohmayer \& Markwardt (1999) have shown that with accurate modeling of the 
frequency drift quality factors as high as $Q \sim 4,000$ are achieved in 
some bursts.  They modelled the frequency drift and showed that a simple 
exponential ``chirp" model of the form $\nu (t) = \nu_0 (1 - \delta_{\nu} 
\exp(-t/\tau) )$, works remarkably well. The resulting quality factors 
derived from the frequency modelling are very nearly consistent with the factors
expected from a perfectly coherent signal of finite duration equal to the
length of the data trains in the bursts. These results argue strongly that
the mechanism which produces the modulations is a highly coherent process,
such as stellar rotation, and that the asymptotic frequencies observed
during bursts represent the spin frequency of the neutron star.

\subsection{The long-term stability of burst oscillation frequencies}

The accretion-induced rate of change of the neutron star spin frequency in a 
LMXB is approximately $1.8 \times 10^{-6}$ Hz yr$^{-1}$ for typical neutron
star and LMXB parameters.  The Doppler shift due to orbital motion of the
binary can produce a frequency shift of magnitude
$ \Delta\nu / \nu = v \sin i /c \approx  2.05 \times 10^{-3}$, again for
canonical LMXB system parameters. This doppler shift easily dominates over 
any possible accretion-induced spin change on orbital to several year 
timescales. Therefore the extent to which the observed burst oscillation 
frequencies are consistent with possible orbital Doppler shifts, but 
otherwise stable over $\approx$ year timescales, provides strong support 
for a highly coherent mechanism which sets the observed frequency. 

At present, the best source available to study the long term stability of burst
oscillations is 4U 1728-34. Strohmayer et al. (1998b) compared the observed
asymptotic frequencies in the decaying tails of bursts separated in time
by $\approx 1.6$ years. They found the burst frequency to be highly stable,
with an estimated time scale to change the oscillation period of 
about 23,000 year. It was also suggested that the stability of the 
asymptotic periods might be used to infer the X-ray mass function of
LMXB by comparing the observed asymptotic period distribution of many
bursts and searching for an orbital Doppler shift.

%

\begin{figure}[htb]
\centerline{\psfig{file=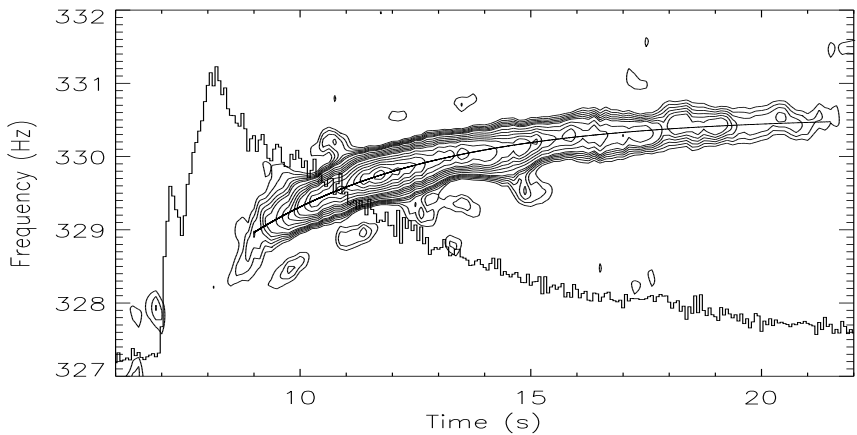,width=5.0in,height=2.5in,clip=}}
\caption{A dynamic power density spectrum of a burst from 4U 1702-429 showing
frequency drift toward an asymptotic limit. The solid curve is a best fit using
an exponential recovery.}
\end{figure}


\section{Burst oscillations as probes of neutron stars}

Detailed studies of the burst oscillation phenomenon hold great promise for 
providing new insights into a variety of physics issues related to the structure
and evolution of neutron stars. In particular, the burst oscillations have given
astronomers their first direct method to investigate the two dimensional nature of
nuclear flame front propagation. In this section I will outline how the burst
oscillations can be used to probe neutron stars.

\subsection{Mass - Radius constraints and the EOS of dense matter}

Using the rotating hot spot model it is possible to determine constraints on the
mass and radius of the neutron star from measurements of the maximum observed
modulation amplitudes during X-ray bursts as well as the harmonic content of the
pulses. The physics that makes such constraints possible is the bending of photon 
trajectories in a strong gravitational field. The strength of the deflection is a
function of the stellar compactness, $GM/c^2 R$, with more compact stars producing
greater deflections and therefore weaker spin modulations. An upper limit on the 
compactness can be set since a star more compact than this limit would not be
able to produce a modulation as large as that observed. Complementary information
comes from the pulse shape, which can be inferred from the strength of harmonics.
Information on both the amplitude and harmonic content can thus be used to bound
the compactness. Detailed modelling, during burst rise for example, can then
be used to determine a confidence region in the mass - radius plane for neutron
stars. Miller \& Lamb (1998) have investigated the amplitude of rotational
modulation pulsations as well as harmonic content assuming emission from a 
point-like hot spot. They also show that knowledge of the angular and spectral 
dependence of the emissivity from the neutron star surface can have important
consequences for the derived constraints. More theoretical as well as data
modelling in this area are required.

\subsection{Doppler shifts and pulse phase spectroscopy}

Stellar rotation will also play a role in the observed properties
of spin modulation pulsations. For example, a 10 km radius neutron star 
spinning at 400 Hz has a surface velocity of $v_{spin}/c \le 2\pi
\nu_{spin} R  \approx 0.084$ at the rotational equator. This motion of
the hot spot produces a Doppler shift of magnitude $\Delta E / E \approx
v_{spin}/c$, thus the observed spectrum is a function of pulse phase 
(see Chen \& Shaham 1989). Measurement of a pulse phase dependent
Doppler shift in the X-ray spectrum would provide additional evidence 
supporting the spin modulation model and also yields a means of 
constraining the neutron star radius, perhaps one of the few direct
methods to infer this quantity for neutron stars.

The rotationally induced velocity also produces a relativistic aberration 
which results in asymmetric pulses, thus the pulse shapes also contain 
information on the spin velocity and therefore the stellar radius 
(Chen \& Shaham 1989). The component of the spin velocity 
along the line of site is proportional to $\cos\theta$, where $\theta$ is the
lattitude of the hotspot measured with respect to the rotational equator. The
modulation amplitude also depends on the lattitude of the hotspot, as spots near
the rotational poles produce smaller amplitudes than those at the equator. Thus
a correlation between the observed oscillation amplitude and the
size of any pulse phase dependent Doppler shift is to be expected. Dectection 
of such a correlation in a sample of bursts would provide strong
confirmation of the rotational modulation hypothesis.

Searches for a Doppler shift signature are just beginning to be carried
out. Studies in single bursts have shown that spectral variations with
pulse phase can be detected (see Strohmayer, Swank, \& Zhang 1998).
The varations with pulse phase show a 4-5 \% modulation of the fitted
black body temperature, consistent with the idea that a temperature
gradient is present on the stellar surface, which when rotated produces
the flux modulations. Ford (1999) has analysed data during a burst from 
Aql X-1 and finds that the softer photons lag higher energy photons in
a manner which is qualitatively similar to that expected from a rotating
hot spot. Strohmayer \& Markwardt (1999) have shown that signals from multiple
bursts can be added in phase by modelling the frequency drifts present 
in individual bursts. This provides a stronger signal with which to test
for Doppler shift effects.  So far, burst oscillation signals from 4U 1702-429 
have been added in phase in an attempt to identify a rotational Doppler
shift. A difficulty in analysing the phase resolved spectra from bursts
is the systematic change in the black body temperature produced as the
surface cools. A simpler measure of the spectral hardness, rather than
the black body temperature, is the mean energy channel of the spectrum.
I computed the distribution of mean channels in the RXTE proportional 
counter array (PCA) as a function of pulse phase using spectra from 
4 different bursts from 4U 1702-429. Figure 4 shows the results. A strong 
modulation of the mean PCA channel is clearly seen. There is a hint of 
an asymmetry in that the leading edge of the pulse appears harder (as 
expected for a rotational Doppler shift) than the trailing edge, but 
the difference does not have a high statistical significance. More data 
will be required to decide the rotational Doppler shift issue.

\begin{figure}[htb]
\centerline{\psfig{file=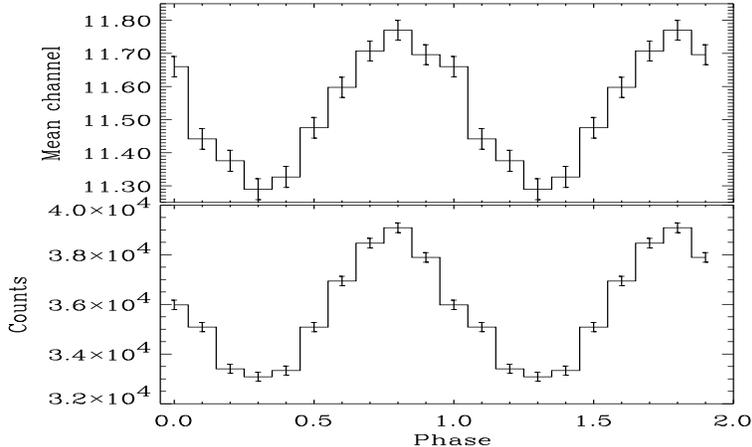,width=4.5in,height=2.5in,clip=}}
\caption{Pulse phase spectral variations in bursts from 4U 1702-429. The
top panel shows the mean PCA energy channel as a function of pulse phase
for 4 bursts which were co-added in phase. The bottom panel shows the 
pulse profile in the 2 - 24 keV band.}
\end{figure}

\subsection{Physics of thermonuclear burning}

The properties of burst oscillations can tell us a great deal about the 
processes of nuclear burning on neutron stars. The amplitude evolution
during the rising phase of bursts contains information on how rapidly
the flame front is propagating. If the anitpodal spot hypothesis to explain
the presence of a subharmonic in 4U 1636-53 is correct, then it has important
implications for the propagation of the instability from one pole to
another in $\approx$ 0.2 s (see Miller 1999). In addition, a two pole flux 
anistropy suggests that the nuclear fuel is likely pooled by some mechanism, 
perhaps associated with the magnetic field of star. Further detections and
study of the subharmonic in 4U 1636-53 could shed more light on these issues.

Until recently, much of the work concerning burst oscillations has concentrated
on studies of the pulsations themselves and their relation to individual bursts. 
With the samples of bursts growing it is now possible to concieve of more global 
studies which correlate the properties of oscillations with other measures of 
these sources, for example, their spectral state and mass accretion rate. 
This will allow researchers to investigate the system parameters which determine 
the likelihood of producing bursts which show oscillations. Such investigations 
will provide insight into how the properties of thermonuclear burning 
(as evidenced in the presence or absence of oscillations) are influenced by other 
properties of the system. Furthermore, we can test if theoretical predictions of 
how the burning should behave are consistent with the hypothesis that 
the oscillations result from rotational modulation of nonuniformities produced by 
thermonuclear burning. Initial work in this regard suggests that bursts which 
occur at higher mass accretion rates show stronger burst oscillations more often
(see Franco et al. 1999). Although preliminary this result appears roughly 
consistent with theoretical descriptions of the thermonuclear burning which 
indicates an evolution from vigorous, rapid (thus uniform) burning at lower 
mass accretion rates (lower persistent count rates) to weaker, slower burning 
(thus more non-uniform) at higher mass accretion rates (see Bildsten 1995, 
for example). 

\section{Remaing puzzles and the future}

Although much of the burst oscillation phenomenology is well described by the
spin modulation hypothesis several important hypotheses need to be confronted
with more detailed theoretical investigations. Perhaps the most interesting
is the mechanism which causes the observed frequency drifts. Expressed as
a phase slip the frequency drifts seen during the longest pulse trains correspond
to about 5 - 10 revolutions around the star. Whether or not a shear layer can
persist that long needs to be further investigated. The recent
observation of a, so far unique, spin down in the decaying tail of a burst from
4U 1636-53 (see Strohmayer 1999), which might be the first detection of the 
spin down caused by thermal expansion of the burning layers, needs to be better
understood in the context of thermonuclear energy release at late times in bursts.
Another perplexing issue is the mechanism which allows flux asymmetries to 
both form and then persist at late times in bursts. 

Although RXTE provided the technical advancements required to discover the 
burst oscillations, it may take future, larger area instruments such as 
Constellation-X or a successor timing mission to RXTE to fully exploit 
their potential for unlocking the remaining secrets of neutron stars.

\end{document}